\documentclass[british,english,prl, singlespace, twocolumn]{revtex4-1}
\usepackage[T1]{fontenc}
\usepackage[latin9]{inputenc}
\usepackage{color}
\usepackage{bm}
\usepackage{amsmath}
\usepackage{amssymb}
\usepackage{graphicx}

\makeatletter
\@ifundefined{definecolor}
 {\usepackage{color}}{}
\makeatother

\makeatother

\usepackage{babel}
\begin{document}
\title{Mesoscopic two-mode entangled and steerable states of 40,000 atoms
in a Bose-Einstein condensate interferometer }
\author{B. Opanchuk$^{1}$, L. Rosales-Zárate$^{2}$, R. Y. Teh$^{1}$, B.
J. Dalton$^{1,3}$, A. Sidorov$^{1},$ P. D. Drummond$^{1,4}$ and
M. D. Reid$^{1,4}$}
\affiliation{$^{1}$Centre for Quantum and Optical Science Swinburne University
of Technology, Melbourne, Australia}
\affiliation{$^{2}$Centro de Investigaciones en Óptica A.C., León, Guanajuato
37150, México}
\affiliation{$^{3}$Centre for Cold Matter, Blackett Laboratory, Imperial College
of Science,Technology and Medicine, London SW7 2BZ, United Kingdom}
\affiliation{$^{4}$Institute of Theoretical Atomic, Molecular and Optical Physics
(ITAMP),Harvard University, Cambridge, Massachusetts, USA}
\begin{abstract}
 \textcolor{black}{Using criteria based on superselection rules,
we analyze the quantum correlations between}\textcolor{black}{\normalsize{}
the two condensate modes of the Bose-Einstein condensate interferometer
of Egorov et al. {[}Phys. Rev. A }\textbf{\textcolor{black}{\normalsize{}84}}\textcolor{black}{\normalsize{},
021605 (2011){]}. In order to determine the two-mode correlations,
we develop a multi-mode theory that describes the dynamics of the
condensate atoms and the thermal fraction through the interferometer
sequence, in agreement with the experimentally measured fringe visibility.
We thus present experimental evidence for two-mode entangled states
genuinely involving 40,000 $^{87}$Rb atoms, and for two-way steerability
between two groups of 20,000 indistinguishable atoms.}{\normalsize\par}
\end{abstract}
\maketitle
\textcolor{black}{In the Einstein, Podolsky and Rosen (EPR) paradox,
a measurement made by an observer at one location can apparently instantaneously
affect the quantum state at another \cite{eprbell}. This effect
was called ``steering'' by Schrodinger \cite{hw-1,Schrodinger-1}.
States that give the correlations of an EPR paradox a}re called steerable,
or EPR steerable if the two locations are spatially separated \cite{hw-1,eric,epr-reid,sjonessteer}.
Although well verified for optical systems \cite{rmpepr,pryde-steer},
it is a challenge to demonstrate EPR steering correlations between
large massive systems.  To resolve paradoxes associated with macroscopic
quantum objects, decoherence theories propose to modify quantum mechanics
by including gravitational effects \cite{furry,diosi-1}, thus distinguishing
between massive and massless systems. For these reasons, the detection
of EPR steering correlations between mesoscopic groups of atoms is
an important milestone. 

There has been success in entangling massive systems \cite{milestonewhy,mech-ent-2,eprenthiedel-1,EntAtoms-1,eprnaturecommun-2,SteerAt-obert-1,treu-matteo-1,bell-kasevich,treutlein-exp-bell}.
\textcolor{black}{Yet entanglement does not imply steering, which
is a stronger form of quantum correlation. Steering is a necessary
(though not sufficient) requirement for all systems that show Bell-nonlocality
\cite{Bell-1}, and is useful for certain quantum information tasks
\cite{steering-app}. Several experimental groups have inferred Bell
or steering correlations between atoms within an atomic ensemble \cite{bell-kasevich,treutlein-exp-bell,eprnaturecommun-2},
and there has been demonstration of Bell correlations  involving
optomechanical oscillators \cite{bell-oscillator}. In a further
step, EPR steering has been observed between spatially separated clouds
of several hundreds of atoms formed from a Bose-Einstein condensate
(BEC) \cite{EntAtoms-1,SteerAt-obert-1,treu-matteo-1}.}

\textcolor{black}{However, there is a difference between states with
many mutually entangled atoms, and states built of multiple smaller
entangled units, such as independent pairs of entangled atoms. This
distinction has motivated experiments that rigorously quantify the
number of atoms genuinely involved in the entangled unit (called the
``depth of entanglement'' \cite{sm-1,depth-steer}), leading to
evidence of states with a few hundred atoms mutually entangled in
a BEC \cite{Gross2010,Philipp2010}, a few thousand in a thermal atomic
ensemble \cite{herald3000}, and up to a few million in a crystal
lattice \cite{frowis-ent}. However, entanglement does not imply
steering, and so far atomic experiments have not addressed the size
of steerable units. Moreover, most experiments have considered entanglement
shared between distinguishable particles. This contrasts with a BEC,
where atoms are indistinguishable particles occupying distinct modes.
Since modes can be separated, demonstrating mode entanglement for
highly occupied modes is promising for obtaining nonlocality between
spatially separated mesoscopic groups of atoms. While mode entanglement
has recently been observed \cite{EntAtoms-1,SteerAt-obert-1,treu-matteo-1},
 the maximum number of atoms involved has been limited to several
hundred.}

\textcolor{black}{In this paper, we present experimental evidence
for atomic two-mode steerable entangled states genuinely involving
40,000 atoms, with 20,000 atoms localised in each condensate mode.
The states are created in a multi-mode $^{87}$Rb BEC Ramsey interferometer
of $\sim55,000$ atoms at a temperature of $\sim37nK$ and prepared
on an atom chip in a magnetic trap \cite{Egorov,Egorov2013}. Steering
is a directional form of entanglement, because one can consider a
nonlocal effect one way, on one system due to measurements on the
other, and vice versa. Two-way steering is required for Bell nonlocality
\cite{hw-1,eric}. Here, we demonstrate that the correlations between
two atomic condensate modes are two-way steerable, thereby inferring
the steerability of 20,000 indistinguishable atoms.}

\textcolor{black}{It is important to clarify the meaning of ``entangled
states genuinely involving $N$ atoms'', in the context of mode entanglement.
The entanglement depth is not simply the number of atoms in the experiment,
nor the number of atoms in the two condensate modes. This is because
the system may be in a mixed state where large numbers of atoms are
in separable (non-entangled) two-mode states. Furthermore, at finite
temperature, a significant number of atoms are lost into thermal modes.
We define the ``mode-entanglement (steering) depth'' as the number
of atoms $N$  in the part of the density operator associated with
two-mode entanglement (steering). Specifically, we will confirm that
the entanglement cannot be explained, if we allow that the number
$N$ is reduced. In this paper, we measure a mode-entanglement and
mode-steering depth of $40,000$ atoms.}

\textbf{\textcolor{black}{Mode versus particle entanglement:}}\textcolor{black}{{}
We may ask how to compare the mode-entanglement depth with the particle-entanglement
depth investigated in previous experiments. Indeed, there has been
controversy about the meaning of particle entanglement when particles
are indistinguishable, and hence not individually localisable, as
in a BEC \cite{PlenioInd,bryan-reviews,wise-vaccaro}. }

\textcolor{black}{To illustrate, consider bosons incident on a Ramsey
BEC interferometer. For two atomic bosonic modes, superselection rules
apply that fix the total particle number $N$ for a pure state \cite{ssr,ssr-molecule,ssrpapers2,bryanlibby,wise-vaccaro}.
The most general pure two-mode state is then of the form ($\mathcal{N}$
is a constant)
\begin{equation}
|\psi_{N}\rangle=\mathcal{N}\sum_{n=0,1,..}^{N}d_{n}\sqrt{\binom{N}{n}}|n\rangle_{a}|N-n\rangle_{b}\label{eq:bs-state}
\end{equation}
where $d_{n}$ are complex amplitudes. Here $|n\rangle_{a}|N-n\rangle_{b}$
denotes $n$ particles in mode $a$ with spin $0$, and $N-n$ particles
in mode $b$ with spin $1.$ The state is mode-entangled for any $d_{n}$,
provided $d_{n}\neq0$ for at least two values of $n$. Following
\cite{PlenioInd}, we write $|n\rangle|N-n\rangle=\frac{1}{\sqrt{\binom{N}{n}}}S|0\rangle_{1}...|0\rangle_{n}|1\rangle_{n+1}...|1\rangle_{N}$
where $S$ denotes symmetrisation of the particle state in first
quantisation \cite{PlenioInd}. If we view the pseudo-labels $1,..,N$
of the symmetrised wave function as corresponding to $N$ distinguishable
particles, then it is straightforward to show that the mode-separable
state $|n\rangle_{a}|N-n\rangle_{b}$ for $n=1,..,N-1$ is both $N$-particle
entangled and $N$-particle steerable \cite{gen-steer,bancalgendient,sm}.
This is also true in general for the mode-entangled state $|\psi_{N}\rangle$,
except for some singular choices such as $d_{n}=1$. Details are
given in the Supplemental Materials \cite{sm,gisin-gen}.}

\textcolor{black}{This provides a link between the mode-entanglement
depth, and the pseudo-label particle-entanglement depth inferred
in earlier experiments \cite{Gross2010,Philipp2010}: A two-mode entangled
state with a mode-entanglement (steering) depth of $N$ is also pseudo-label
particle-entangled (steerable) with depth $N$, except in the singular
cases. In those cases, once we determine the value $N$ of the mode-entanglement
(steering) depth, a state with pseudo-label $N$-particle entanglement
(steering) can be prepared by a local operation that projects onto
a definite local mode number $n$ \cite{wise-vaccaro}. Although
this $N$-particle entanglement is without operational meaning (since
pseudo-labelled systems are not independently measurable \cite{partition-unphysical,PlenioInd,wise-vaccaro,bryan-reviews}),
such particle entanglement can be transformed into multi-partite
mode entanglement by expanding and splitting the BEC \cite{PlenioInd,SteerAt-obert-1,treu-matteo-1}.
An $N$-partite entanglement can only be realised however, once each
atom is localisable.}

\textcolor{black}{The observation of any degree of spin squeezing
is sufficient to imply a (pseudo-label) $N$-particle entanglement,
once the mode-entanglement depth $N$ has been determined. This follows
because the particles are indistinguishable \cite{sm}.  In our
experiment, a spin squeezed state $|\psi_{N}\rangle$ is predicted
as the atoms evolve \cite{sm}, but we do not measure this effect.
 In fact, $|\psi_{N}\rangle$ with $d_{n}=1$ is an approximate model
for the state generated. While such a state is separable with respect
to the pseudo-labels,  this has limited meaning because the subsystems
are not distinguishable. In particular, if we envisage preparing each
atom independently in its own mode to ensure localisation (with each
split equally between $a$ and $b$), then, because the atoms are
indistinguishable, the resulting symmetrised wave function in the
coordinate representation is not factorisable with respect to the
$N$ modes \cite{wise-vaccaro}. Most relevant is that the operational
entanglement between $a$ and $b$ as measured by the entropy of entanglement
is of order $N$ (or $\frac{1}{2}\log N$ for $|\psi_{N}\rangle$
with $d_{n}=1$), illustrating a cooperative efffect due to all $N$
bosons (refer to the Supplemental Materials for details
of the proofs) \cite{wise-vaccaro,sm}.}

\textbf{\textcolor{black}{Steering:}}\textcolor{black}{{} We begin by
defining the concept of steering for two systems $a$ and $b$ \cite{hw-1}.
Where each system is a single mode, we introduce boson creation and
destruction operators $\hat{a}^{\dagger}$, $\hat{a}$, $\hat{b}^{\dagger}$,
$\hat{b}$ for $a$ and $b$ respectively. The two systems are entangled
if the quantum density operator $\rho$ of the composite system cannot
be described according to a separable model $\rho=\sum_{R}P_{R}\rho_{a}^{R}\rho_{b}^{R}$.
Here, $\rho_{a}^{R}$ and $\rho_{b}^{R}$ are density operators for
$a$ and $b$, and $P_{R}$ are probabilities satisfying $\sum_{R}P_{R}=1$
and $P_{R}>0$. If the modes are at different locations, EPR steering
of $b$ by $a$ is demonstrated if there is a failure of all local
hidden state models, where the averages for locally measured observables
$X_{a}$ and $X_{b}$ are given as \cite{hw-1,sjonessteer} 
\begin{equation}
\langle X_{b}X_{a}\rangle=\int_{\lambda}P(\lambda)d\lambda\langle X_{b}\rangle_{\rho,\lambda}\langle X_{a}\rangle_{\lambda}\label{eq:lhs}
\end{equation}
The states symbolized by $\lambda$ are the hidden variable states
introduced in Bell's local hidden variable models, with probability
density $P(\lambda)$ satisfying $\int_{\lambda}P(\lambda)d\lambda=1$.
$\langle X_{a}\rangle_{\lambda}$ is the average outcome of $X_{a}$
given the system is in the state $\lambda$. The $\rho$ subscript
denotes that the average $\langle X_{b}\rangle_{\rho,\lambda}$ is
generated from a local quantum state with quantum density matrix $\rho_{b}^{\lambda}$.}

\textbf{\textcolor{black}{Entangled modes of an interferometer: }}\textcolor{black}{The
entangled states reported in this paper can be understood using a
simple model of a Mach-Zehnder interferometer (Figure 1). Consider
two field modes impinging on a 50/50 beam splitter $BS1$. The input
state for mode $a$ is a Fock number state $|N\rangle_{a}$ describing
$N$ bosons. The input to $b$ is the vacuum state $|0\rangle_{b}$.
The output of the beam splitter is the two-mode entangled state $|\psi_{N}\rangle$
(\ref{eq:bs-state}) where $d_{n}=1$, $\mathcal{N}=1/\sqrt{2^{N}}$
\cite{kim-bs}. Equivalent predictions are given for a BEC atom Ramsey
interferometer. The incident mode $|N\rangle$ represents $N$ atoms
of a single-component BEC prepared in an atomic hyperfine level $|1\rangle=|F=1,$$m_{F}=-1\rangle$
(spin $0$). A  $\pi/2$ microwave pulse creates a two-component
BEC associated with two hyperfine levels $|1\rangle$ and $|2\rangle=|F=1,m_{F}=1\rangle$
(spin $1$). This produces the action of the beam splitter $BS1$,
creating the mode-entangled state $|\psi_{N}\rangle$. The components
$|1\rangle$ and $|2\rangle$ correspond to well-defined spatial condensate
modes $a$ and $b$. The nonlinearity $\chi$ of the BEC gives rise
to enhanced entanglement, and an $N$-particle entanglement with
respect to particle pseudo-labels \cite{esteve,hesteer,depth-steer,Philipp2010,sm-1,Gross2010,bogdan-dynamics,yun-li-2,yunli-2}.
After an evolution time $T$, a second interrogating microwave pulse
is applied with a phase lag $\varphi$, producing the action of a
second beam splitter. Immediately after, the atoms are released and
the two-component population difference $\langle\hat{N}_{-}\rangle$
measured by atom imaging. The size of the moment $\langle\hat{a}^{\dagger}\hat{b}\rangle$
can be used to detect the entanglement between the modes \cite{hesteer,hilzub,caval-crit}.
 It is difficult however to use existing criteria \cite{hesteer,hilzub,caval-crit,depth-steer},
due to the difficulty of preparing a state with an exact atom number
$N$.}

\begin{figure}[t]
\includegraphics[width=0.8\columnwidth]{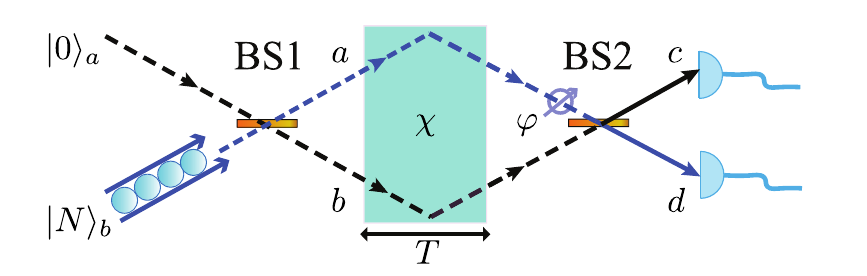}

\caption{\textcolor{black}{Schematic of a two-mode Mach-Zehnder interferometer.
Entangled modes $a$ and $b$ are prepared by means of a number state
$|N\rangle_{a}$ incident on the first beam splitter BS1. The entanglement
may be enhanced by a nonlinear interaction $\chi$ acting for a time
$T$. }\label{fig:Data-for-EPR}}
\end{figure}

\textbf{\emph{Super-selection rules and criteria for steering:}} Two-mode
entanglement and steering can regardless be inferred if the bosons
are \emph{atoms}, using an alternative two-mode criterion 
\begin{equation}
\langle\hat{a}^{\dagger}\hat{b}\rangle\neq0\label{eq:crit}
\end{equation}
sufficient to confirm both entanglement and a two-way steering between
modes $a$ and $b$. The criterion is based on superselection rules
that forbid superpositions of eigenstates of different single-mode
atom number \cite{ssr,ssr-molecule,ssrpapers2,bryanlibby,bryan-reviews,bry-arxiv-steer-ssr,wise-vaccaro}.
Following Refs. \cite{bryanlibby,bry-arxiv-steer-ssr}, we
give proof of the condition (\ref{eq:crit}). The density operator
for any separable state can be written $\rho=\sum_{R}P_{R}\rho_{a}^{R}\rho_{b}^{R}$.
According to the superselection rule, the single-mode atom coherences
$\langle\hat{a}\rangle_{R}$ and $\langle\hat{b}\rangle_{R}$ vanish,
for any local single mode quantum states $\rho_{a}^{R}$ and $\rho_{b}^{R}$.
Thus, the separable model implies $\langle\hat{a}^{\dagger}\hat{b}\rangle=\sum_{R}P_{R}(\langle\hat{a}\rangle_{R})\langle\hat{b}\rangle_{R}=0$,
as does the local hidden state model (\ref{eq:lhs}). Unless we allow
that the individual modes violate the superselection rule, the observation
of $\langle\hat{a}^{\dagger}\hat{b}\rangle\neq0$ is sufficient to
confirm entanglement, and a ``two-way'' steering ($b$ by $a$,
and $a$ by $b$) between the modes.

Ultimately, we envisage detecting $\langle\hat{a}^{\dagger}\hat{b}\rangle\neq0$
using localized measurements on each of the modes, in the spirit of
the Einstein-Podolsky-Rosen argument \cite{eprbell}. This is possible
using quadrature phase amplitudes $\hat{X}_{a}=\hat{a}+\hat{a}^{\dagger}$,
$\hat{P}_{a}=(\hat{a}-\hat{a}^{\dagger})/i$ and $\hat{X}_{b}=\hat{b}+\hat{b}^{\dagger}$,
$\hat{P}_{b}=(\hat{b}-\hat{b}^{\dagger})/i$, since one can expand
\textbf{$\langle\hat{a}^{\dagger}\hat{b}\rangle=(\langle\hat{X}_{a}\hat{X}_{b}\rangle+\langle\hat{P}_{a}\hat{P}_{b}\rangle-i\langle\hat{P}_{a}\hat{X}_{b}\rangle+i\langle\hat{X}_{a}\hat{P}_{b}\rangle)/4$}
\cite{eprenthiedel-1,eprnaturecommun-2,SteerAt-obert-1}. As a preliminary
step to such an observation, we show that $\langle\hat{a}^{\dagger}\hat{b}\rangle\neq0$
can be inferred, based on interferometry. Introducing a phase shift
$\varphi$, the two-mode outputs of the interferometer are described
by operators $\hat{c}=(\hat{a}-\hat{b}\exp^{i\varphi})/\sqrt{2}$,
$\hat{d}=(\hat{a}+\hat{b}\exp^{i\varphi})/\sqrt{2}$. Defining $\hat{N}_{\pm}=\hat{d}^{\dagger}\hat{d}\pm\hat{c}^{\dagger}\hat{c}$
and assuming $N_{+}$ to be fixed, the normalized average population
difference $P_{z}=N_{-}/N_{+}$ at the output is $P_{z}=2(Re\langle\hat{a}^{\dagger}\hat{b}\rangle\cos\varphi-Im\langle\hat{a}^{\dagger}\hat{b}\rangle\sin\varphi)/N_{+}$
($N_{\pm}$ are the outcomes of $\hat{N}_{\pm}$). By adjusting $\varphi$,
$|\langle\hat{a}^{\dagger}\hat{b}\rangle|$ can be inferred from the
interference fringe amplitude $\nu$, ($\nu=2|\langle\hat{a}^{\dagger}\hat{b}\rangle|$)
\cite{Egorov,Egorov2013,sm}. The observed fringe pattern for the
BEC interferometer is given in Figure 2. While $T$ and $\varphi$
can be controlled experimentally, there are run-to-run fluctuations
in the total atom number $N_{+}$. The criterion $\langle\hat{a}^{\dagger}\hat{b}\rangle\neq0$
is however valid for all mixed two-mode states and hence applies to
fluctuating number inputs.
\begin{figure}[b]

\includegraphics[width=0.8\columnwidth]{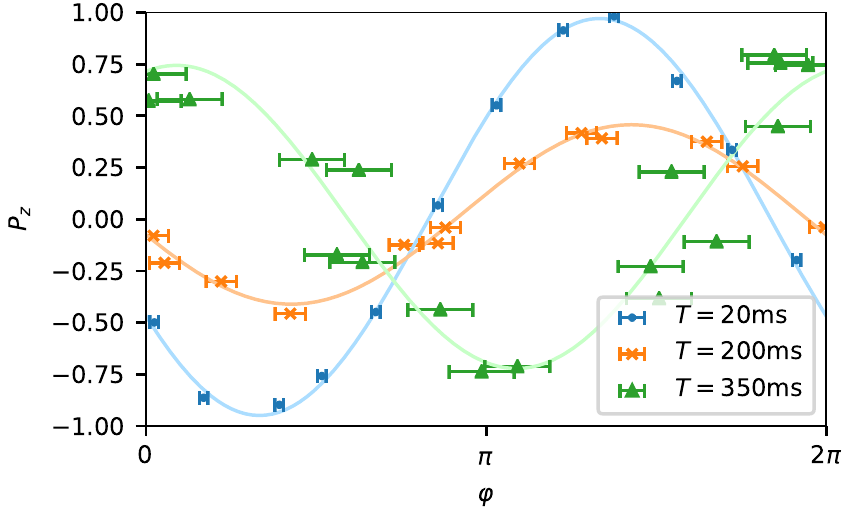}

\caption{\textcolor{black}{}\textbf{\textcolor{black}{}}\textcolor{black}{The
plot shows the experimentally observed interference at $T=20ms$,
$T=200ms$, $T=350ms$.}\textcolor{red}{{} }\textcolor{black}{The $P_{z}=N_{-}/N_{+}$
is the normalized population difference after a correction $\phi(N_{+},T)$
is added as explained in Refs. \cite{Egorov} to account for the effect
of fluctuating population number $N_{+}$. Here $N_{+}\sim10^{4}$
atoms. The solid line is the best fit to the data. The observed fringe
amplitude is larger than predicted by a two-mode model, due to the
presence of thermal atoms.}}
\end{figure}

\textbf{\emph{Depth of steering:}} We next address how to determine
the number of atoms in the steerable unit $-$ the ``mode-steering
depth $n_{st}$'' \cite{depth-steer}. This is not given by the mean
number $\langle N\rangle$ of particles because the system is generally
a\emph{ }mixture of pure states $\left\vert \psi_{R}\right\rangle $,
according to a density operator $\rho=\sum_{R}P_{R}\left\vert \psi_{R}\right\rangle \left\langle \psi_{R}\right\vert $
($\sum_{R}P_{R}=1,$$P_{R}>0$). While laboratory preparations of
a BEC are near-pure states, a mixed state analysis is required because
of finite temperatures and fluctuations in the atom number. Each pure
state $|\psi_{R}\rangle$ has a fixed number of atoms (according to
superselection rules) that we denote by $n_{R}=\langle\psi_{R}|N|\psi_{R}\rangle$.
However, not all the $|\psi_{R}\rangle$ need be steerable. Defining
$n_{st}$ to be the maximum value of $n_{R}$ taken over all the $|\psi_{R}\rangle$
that are steerable, we prove in the Supplementary Materials that \cite{sm}
\begin{equation}
n_{st}\geq2|\langle\hat{a}^{\dagger}\hat{b}\rangle|\label{eq:condition-number}
\end{equation}

We might apply the criterion to known experimental systems e.g. the
creation of two steerable modes is possible for a BEC in a double-well
potential \cite{esteve,hesteer}. \textcolor{red}{ }In order to
properly quantify the two-mode correlation $\langle\hat{a}^{\dagger}\hat{b}\rangle$
for larger BECs, however, a full multi-mode model is necessary. \textcolor{black}{This
is particularly true for higher temperatures, and is necessary because
the extra modes involving thermal atoms contribute to the measured
fringe contrast. Some atom interferometers have large fringe visibilities
and yet comprise multiple thermally-excited modes, with a small occupation
of each mode (see Refs. \cite{bell-kasevich,anu-exp-1,deutsch-thermal-int}).}

\textbf{\emph{Multi-mode BEC interferometer:}} To infer a steerable
state of thousands of atoms, we calculate the condensate fractions
in the BEC interferometer using the Onsager-Penrose criterion \cite{Penrose1956}.
The quantum dynamics are evaluated using a multi-mode field-theoretic
phase-space method based on the Wigner function \cite{Egorov,bogepl}.
The effective Hamiltonian for the two-mode condensate system is \cite{Widera2006,Mertes2007,yun-li-2,yunli-2,bogepl}:\textcolor{red}{{} }

\begin{equation}
\hat{H}=\int d^{3}\bm{x}\sum_{k,j=1}^{2}\left\{ \hat{\Psi}_{i}^{\dagger}K_{ij}\hat{\Psi}_{j}+\frac{g_{ij}}{2}\hat{\Psi}_{i}^{\dagger}\hat{\Psi}_{j}^{\dagger}\hat{\Psi}_{i}\hat{\Psi}_{j}\right\} ,
\end{equation}
where $\hat{\Psi}_{j}$ describes a bosonic quantum field operator
with internal spin orientation labelled $j=1,2$ for the two levels
$|1\rangle$, $|2\rangle$ (corresponding to spin states $|0\rangle$
and $|1\rangle)$. Here $g_{jk}=4\pi\hbar^{2}a_{jk}/m$ gives the
S-wave scattering interaction strength, and the single-particle Hamiltonian
operator is $K_{ij}=\left(-\hbar^{2}\nabla^{2}/2m+V\left(\bm{x}\right)\right)\delta_{ij}+\hbar\Omega_{ij}\left(t\right)$.
The important terms are the atomic mass $m$, a trap potential $V\left(\bm{x}\right)=m\sum_{j}\omega_{j}^{2}x_{j}^{2}/2$
and an inter-level Rabi cycling matrix $\Omega_{ij}$.  Previous
work calculating a static condensate fraction used both the semiclassical
Hartree-Fock (SHF) approximations and Monte-Carlo methods~\cite{Holzmann},
showing excellent agreement of these methods in thermal equilibrium,
far from the critical point. This has also been accurately verified
experimentally \cite{Gerbier2004-experimental-study}.

\begin{figure}[t]
\includegraphics[width=0.8\columnwidth]{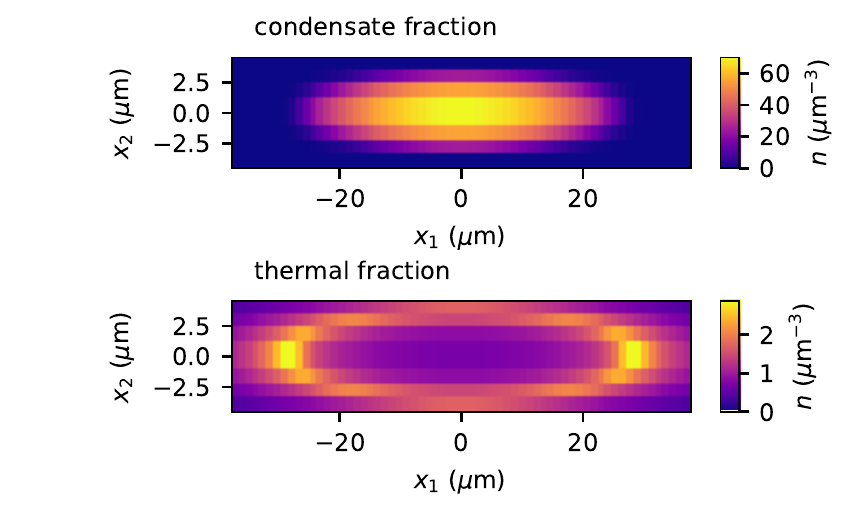}

\caption{\textcolor{black}{The three-dimensional model gives details of the
initial condensate fraction along the axial coordinates of the interferometer.
}The slices shown are taken along the long axis of the trap, and give
the densities of the initial condensate (lower) and thermal (top)
fractions. The total atom population (thermal and BEC fractions) is
$55000$. The initial total condensate population is $N=48325$.}
\end{figure}

\begin{figure}[b]
\includegraphics[width=0.8\columnwidth]{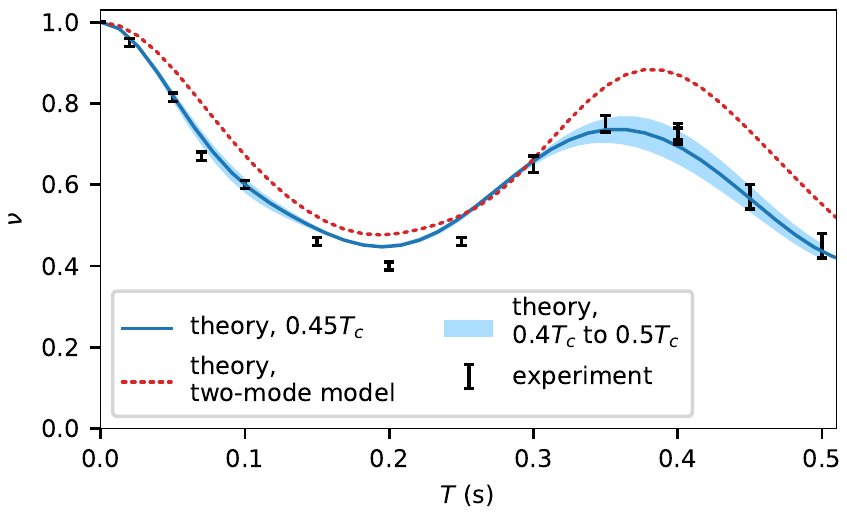}

\caption{\textcolor{black}{The black points give the experimentally observed
fringe contrast $\nu$ versus the evolution time $T$. The blue curve
is the fringe visibility predicted by the multimode theoretical model.
The initial temperature is $T_{BEC}=0.45T_{c}$ where $T_{c}$ is
the critical temperature at which the atoms form a condensate. The
red line shows visibilities obtained with the assumption of a zero-temperature
two-mode model. The thickness of the blue curve corresponds to the
range produced from the values $T=0.4T_{c}$ and $T=0.5T_{c}$ to
demonstrate the error due to the initial temperature estimate.}}
\end{figure}

To obtain the initial density matrix $\rho_{initial}$ we use the
SHF method. This describes the initial finite temperature ensemble
of a three-dimensional, trapped BEC as a coherent condensate $\phi_{j}(\bm{x})$
surrounded by a thermal cloud with occupation $n_{j}^{(T)}(\bm{x})$
(Fig. 3). The thermal fraction density $n^{(T)}$ and the condensate
fraction density $n^{(c)}\equiv|\phi|^{2}$ for the first component
are found self-consistently. 

Since the state is no longer in thermal equilibrium after the action
of the first beam splitter, the condensate evolves dynamically until
rotated back to finish the experiment. To solve the evolution, it
is necessary to go beyond Hartree-Fock approximations. Due to thermal
atoms which form a halo around the central condensate at finite temperature
(Fig. 3), there are large numbers of field modes participating both
in the initial quantum ensemble and its evolution, as well as atomic
losses. To model these effects, the quantum field dynamics is mapped
into a phase-space using a master equation and truncated Wigner approximation
valid at large atom number $N$~\cite{Drummond1993,Steele1998,Opanchuk2013-functional}.
Each quantum field $\hat{\Psi}_{j}$ is transformed into an equivalent
ensemble of complex stochastic fields $\psi_{j},$ that obey a stochastic
partial differential equation which is numerically solved.
\begin{figure}[t]
\includegraphics[width=0.8\columnwidth]{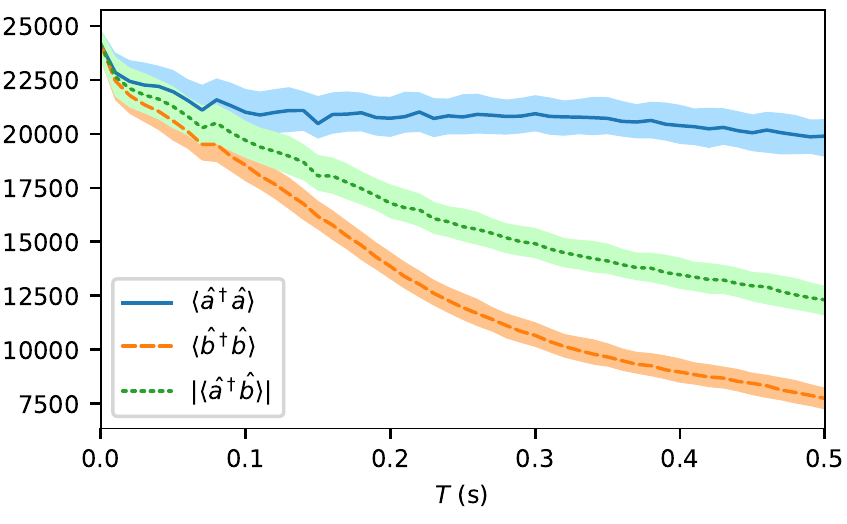}

\caption{\textcolor{black}{The curves show the number of atoms $\langle\hat{a}^{\dagger}\hat{a}\rangle$
and $\langle\hat{b}^{\dagger}\hat{b}\rangle$ in the condensate modes
$a$ and $b$ of each atomic component, and the two-mode moment $\langle\hat{a}^{\dagger}\hat{b}\rangle$,
inferred from the model. The curves in each pair correspond to the
values $T=0.4T_{c}$ and $T=0.5T_{c}$ to demonstrate the error due
to the initial temperature estimate.}}
\end{figure}

The initial condition is assumed to be a grand canonical ensemble
in one of the two components, with an approximately Poissonian number
distribution. For comparison purposes, we consider two initial states.
In one of the plotted lines of Figure 4, we use a coherent state with
average density equal to the solution of the mean-field Gross-Pitaevskii
equation. 

 The absolute temperature is obtained by fitting to the observed
fringe visibility (Fig. 3). This yields an upper bound to the temperature,
expressed as a fraction of the ideal gas critical temperature $T_{c}$
at the same atom number, since there are other technical noise effects
that may slightly degrade the visibility as well. We find\foreignlanguage{british}{
$T_{BEC}=(0.45\pm0.05)T_{c}\approx37nK$}, where $T_{c}=\left(\hbar\bar{\omega}/k_{B}\right)\left(N/\zeta(3)\right)^{1/3}\approx83\,\mathrm{nK}$
is the nominal critical temperature below which the BEC starts to
form for a noninteracting gas with mean trap frequency $\bar{\omega}=\left(\omega_{1}\omega_{2}\omega_{3}\right)^{1/3}$.

Our model includes the spatial evolution of the wave functions, and
thus accounts for the experimentally observed oscillation of the fringe
visibility as a function of $T$ (Fig. 4). One mode decays more quickly
due to inelastic scattering (Fig. 5). Interatomic repulsion is larger
for different states, leading to the fringe visibility oscillation
as the two modes move apart - thus reducing fringe contrast - and
then back together, due to the trap potential.

The data shown in Figure 4 gives the value of the two-mode moment
as $|\langle\hat{a}^{\dagger}\hat{b}\rangle|=20,000$. Using the bound
$n_{st}\geq2|\langle\hat{a}^{\dagger}\hat{b}\rangle|$, this implies
a depth of mode steering (and entanglement) of at least $n_{st}=40,000$
atoms. Moreover, using the criterion based on $|\langle\hat{a}^{\dagger}\hat{b}\rangle|$,
we see the steering is ``two-way''. We thus demonstrate the steering
of an atomic system of at least 20,000 atoms, by another.

\textcolor{black}{Finally, we note that in the experiment the two
condensate modes spatially separate at time T $\sim$ 0.2s, due to
the excitation of collective oscillations \cite{Egorov,Egorov2013},
before being brought back together with minimal loss of quantum coherence
and a revival of interference contrast, at T $\sim$ 0.35s. A similar
spontaneous separation of two-mode functions and associated revival
of the Ramsey contrast has been recently reported in a BEC interferometer
with 5000 atoms \cite{laudat-2018-spinsq}, to give evidence of spin
squeezing. Assuming mechanisms for decoherence with spatial separation
of the modes would likely destroy EPR correlations irreversibly, these
results are promising that mesoscopic EPR steering correlations will
be detected at full spatial separation of the modes. Magnetic fields
could be used to achieve greater spatial separations,  since the
modes correspond to different spin states.}
\begin{acknowledgments}
This research has been supported by the Australian Research Council
Discovery Project Grants schemes under Grant DP140104584 and DP180102470.
This work was performed in part at Aspen Center for Physics, which
is supported by National Science Foundation grant PHY-1607611. BD
thanks the Centre for Cold Matter, Imperial College for hospitality
during this research. We thank V. Ivannikov and M. Egorov for assistance
with obtaining experimental data. 
\end{acknowledgments}


\begin{thebibliography}{10}
\bibitem{eprbell}A. Einstein, B. Podolsky and N. Rosen, Phys. Rev.
\textbf{47}, 777 (1935).

\bibitem{Schrodinger-1}E. Schrödinger, Proc. Camb. Phil. Soc. \textbf{31},
555 (1935).

\bibitem{hw-1}H. M. Wiseman, S. J. Jones and A. C. Doherty, \textit{\emph{Phys.
Rev. Lett. }}\textbf{98}, 140402 (2007).

\bibitem{sjonessteer}S. J. Jones\textcolor{black}{, H. M. Wiseman
and A. Doherty,} \textit{\emph{Phys. Rev. A}} \textbf{76}, 052116
(2007).

\bibitem{eric}E. G. Cavalcanti, S. J. Jones, H. M. Wiseman and M.
D. Reid, Phys. Rev. A\textbf{. 80}, 032112 (2009).

\bibitem{epr-reid}M. D. Reid, Phys. Rev. A \textbf{40}, 913 (1989).

\bibitem{rmpepr}M. D. Reid, P. D. Drummond, W. P. Bowen, E. G. Cavalcanti,
P. K. Lam, H. A. Bachor, U. L. Andersen and G. Leuchs. Rev. Mod.
Phys. \textbf{81}, 1727 (2009).

\bibitem{pryde-steer}D. J. Saunders, S. J. Jones, H. M. Wiseman and
G. J. Pryde,  Nature Physics \textbf{6}, 845 (2010).

\bibitem{furry}W. H. Furry,  Phys. Rev. \textbf{49}, 393 (1936).

\bibitem{diosi-1}A. Bassi, K. Lochan, S. Satin, T. P. Singh and H.
Ulbricht,  Rev. Mod. Phys. \textbf{85}, 471 (2013).

\bibitem{eprnaturecommun-2}J. Peise, I. Kruse, K. Lange, B. Lücke,
L. Pezzè, J. Arlt, W. Ertmer, K. Hammerer, L. Santos, A. Smerzi and
C. Klempt,  Nature Communications \textbf{6}, 8984 (2015).

\bibitem{treutlein-exp-bell}R. Schmied, J.-D. Bancal, B. Allard,
M. Fadel, V. Scarani, P. Treutlein and N. Sangouard, Science \textbf{352},
441 (2016).

\bibitem{bell-kasevich}\textcolor{black}{N. J. Engelsen, R. Krishnakumar,
O. Hosten and M. A. Kasevich,  Phys. Rev. Lett.}\textbf{\textcolor{black}{{}
118}}\textcolor{black}{, 140401 (2017).}

\bibitem{SteerAt-obert-1}P. Kunkel, M. Prüfer, H. Strobel, D. Linnemann,
A. Frölian, T. Gasenzer, M. Gärttner and M. K. Oberthaler, Science
\textbf{360}, 413 (2018).

\bibitem{treu-matteo-1}M. Fadel, T. Zibold, B. Décamps and P. Treutlein,
Science\textbf{ 360}, 409 (2018).

\bibitem{milestonewhy}\textcolor{black}{R. Riedinger, A. Wallucks,
I. Marinkovi\'{c}, C. Löschnauer, M. Aspelmeyer, S. Hong, and S. Gröblacher,
Nature }\textbf{\textcolor{black}{556}}\textcolor{black}{, 473 (2018).}

\bibitem{mech-ent-2}\textcolor{black}{C. F. Ockeloen-Korppi, E. Damskägg,
J.-M. Pirkkalainen, M. Asjad, A. A. Clerk, F. Massel, M. J. Woolley,
and M. A. Sillanpää,  Nature }\textbf{\textcolor{black}{556}}\textcolor{black}{,
478 (2018).}

\bibitem{eprenthiedel-1}C. Gross, H. Strobel, E. Nicklas, T. Zibold,
N. Bar-Gill, G. Kurizki and M. K. Oberthaler, ANature \textbf{480},
219 (2011).

\bibitem{EntAtoms-1} K. Lange, J. Peise, B. Lücke, I. Kruse, G. Vitagliano,
I. Apellaniz, M. Kleinmann, G. Tóth and C. Klempt, Science 3\textbf{60},
416 (2018).

\bibitem{Bell-1}J. S. Bell, . Physics \textbf{1}, 195 (1964).

\bibitem{steering-app}C. Branciard, E. G. Cavalcanti, S. P. Walborn,
V. Scarani, and H. M. Wiseman, Phys. Rev. A \textbf{85}, 010301(R)
(2012). Q. He, L. Rosales-Zarate, G. Adesso, and M Reid, Phys. Rev.
Lett. \textbf{115}, 180502 (2015). M. D Reid, Phys. Rev. A \textbf{88},
062338 (2013). T. Wasak and J. Chwede\'{n}czuk, Phys. Rev. Lett.
\textbf{120}, 140406 (2018).

\bibitem{bell-oscillator}I. Marinkovi\'{c}, A. Wallucks, R. Riedinger,
S. Hong, M. Aspelmeyer, and S. Gröblacher, Phys. Rev. Lett. \textbf{121},
220404 (2018).

\bibitem{sm-1}A. S. Sørensen and K. Mølmer,  \textit{\emph{Phys.
Rev. Lett.}}\emph{ }\textbf{86}, 4431 (2001).

\bibitem{depth-steer}L. Rosales-Zárate, B. Dalton, and M. Reid, \textcolor{black}{Phys.
Rev. A }\textbf{\textcolor{black}{98}}\textcolor{black}{, 022120 (2018).}

\bibitem{Gross2010}C. Gross, T. Zibold, E. Nicklas, J. Esteve, and
M. K. Oberthaler,  Nature (London) \textbf{464}, 1165 (2010).

\bibitem{Philipp2010}M. F. Riedel, P. Böhi, Y. Li, T. W. Hänsch,
A. Sinatra, and P. Treutlein, Nature (London) \textbf{464}, 1170
(2010).

\bibitem{herald3000}R. McConnell, H. Zhang, J. Hu, S. Cuk, and V.
Vuletic,  Nature \textbf{519}, 439 (2015).

\bibitem{frowis-ent}F. Fröwis, P. C. Strassmann, A. Tiranov, C. Gut,
J. Lavoie, N. Brunner, F. Bussières, M. Afzelius and N. Gisin,  Nature
Communications \textbf{8}, 907 (2017).

\bibitem{Egorov}M. Egorov, R. P. Anderson, V. Ivannikov, B. Opanchuk,
P. Drummond, B. V. Hall and A. I. Sidorov, Phys. Rev. A \textbf{84},
021605 (2011).

\bibitem{Egorov2013}M. Egorov, B. Opanchuk, P. D. Drummond, B. V.
Hall, P. Hannaford, and A. I. Sidorov, Phys. Rev. A \textbf{87},
053614 (2013).

\bibitem{wise-vaccaro}H. M. Wiseman and J. A. Vaccaro,\textcolor{black}{{}
 }Phys. Rev. Lett. \textbf{91}, 097902 (2003).

\bibitem{PlenioInd}N. Killoran, M. Cramer, and M. B. Plenio,  Phys.
Rev. Lett. \textbf{112}, 150501 (2014).

\bibitem{bryan-reviews}B. J. Dalton, J. Goold, B. M. Garraway, and
M. D. Reid, Physica Scripta \textbf{92}, 023004 (2017). B. J. Dalton,
J. Goold, B. M. Garraway, and M. D. Reid  (2017); ibid, Physica
Scripta \textbf{92}, 023004 (2017).

\bibitem{bryanlibby} B. J. Dalton, L. Heaney, J. Goold, B. M. Garraway,
and Th. Busch,  New J. Phys., \textbf{16}, 013026 (2014).

\bibitem{ssr}G. C. Wick, A. S. Wightman, and E. P. Wigner,  Phys.
Rev. \textbf{88}, 101 (1952).

\bibitem{ssrpapers2}S. Bartlett, T. Rudolph, and R. Spekkens,  Rev.
Mod. Phys. \textbf{79}, 555 (2007).

\bibitem{ssr-molecule}M. R. Dowling, S. D. Bartlett, T. Rudolph,
and R. W. Spekkens.  Phys. Rev. A \textbf{74}, 052113 (2006).

\bibitem{bancalgendient}J. Bancal, N. Gisin, Y. C. Liang, and S.
Pironio,  Phys. Rev. Lett. \textbf{106,} 250404 (2011).

\bibitem{gen-steer}Q. Y. He and M. D. Reid,  Phys Rev Lett. \textbf{111},
250403 (2013).

\bibitem{sm}See Supplemental Materials for the proof of the pseudo-label
$N$-partite entanglement and $N$-partite steering of the two-mode
state (1), the derivation of the entropy of entanglement for the two-mode
state, the derivation of the criterion for depth of steering, a description
of the BEC interferometer, and of the three-dimensional multimode
model.

\bibitem{gisin-gen}S. Yu, Q. Cheng, C. Zhang, C. H. Lai, and C. H.
Oh, Phys. Rev. Lett. \textbf{109},120402 (2012).

\bibitem{partition-unphysical}P. Zanardi,  Phys. Rev. A \textbf{65,}
042101 (2002). H. Barnum, E. Knill, G. Ortiz, R. Somma, and L. Viola,
 Phys. Rev. Lett. \textbf{92}, 107902 (2004). 

\bibitem{kim-bs}M. S. Kim, W. Son, V. Bu\u{ }zek, and P. L. Knight,
 Phys. Rev. A\textbf{ 65}, 032323 (2002).

\bibitem{yun-li-2}Y. Li, Y. Castin, and A. Sinatra, Phys. Rev. Lett.
\textbf{100}, 210401 (2008). 

\bibitem{yunli-2}Y. Li, P. Treutlein, J. Reichel, and A. Sinatra.
Eur. Phys. J. B \textbf{68}, 365 (2009).

\bibitem{esteve}J. Estève, C. Gross, A. Weller, S. Giovanazzi, and
M. K. Oberthaler,  Nature \textbf{455}, 1216 (2008).

\bibitem{hesteer} Q. Y. He, P. D. Drummond, M. K. Olsen, and M. D.
Reid\textit{\emph{, }} \textit{\emph{Phys. Rev. A}} \textbf{86,}
023626 (2012).

\bibitem{bogdan-dynamics}B. Opanchuk, Q. Y. He, M. D. Reid, and P.
D. Drummond, Phys. Rev. A \textbf{86}, 023625 (2012).

\bibitem{hilzub}M. Hillery and M. S. Zubairy,  Phys. Rev. Lett.
\textbf{96}, 050503 (2006).

\bibitem{caval-crit}E. G. Cavalcanti, Q. Y. He, M. D. Reid, and H.
M. Wiseman,  Phys. Rev. A\textbf{ 84}, 032115 (2011).

\bibitem{bry-arxiv-steer-ssr}B. J. Dalton, B. Garraway, and M. D.
Reid,\textcolor{red}{{} }\textcolor{black}{quant-ph arXiv: 1611.09101.}

\bibitem{deutsch-thermal-int}C. Deutsch, F. Ramirez-Martinez, C.
Lacroûte, F. Reinhard, T. Schneider, J. N. Fuchs, F. Piéchon, F. Laloë,
J. Reichel, and P. Rosenbusch, Phys. Rev. Lett. \textbf{105}, 020401(2010).

\bibitem{anu-exp-1}K. S. Hardman, P. B. Wigley, P. J. Everitt, P.
Manju, C. C. N. Kuhn, and N. P. Robins, Opt. Lett. \textbf{41}, 2505
(2016).

\bibitem{Penrose1956} O. Penrose and L. Onsager, Phys. Rev. \textbf{104},
576 (1956).

\bibitem{bogepl}B. Opanchuk, M. Egorov, S. Hoffmann, A. Sidorov,
and P. Drummond,  Europhys. Lett., \textbf{97}, 50003 (2012).

\bibitem{Widera2006}A. Widera, F. Gerbier, S. Fölling, T. Gericke,
O. Mandel, and I. Bloch, PNew J. Phys. \textbf{8}, 152 (2006).

\bibitem{Mertes2007}K. M. Mertes, J. W. Merrill, R. Carretero-González,
D. J. Frantzeskakis, P. G. Kevrekidis, and D. S. Hall, Phys. Rev.
Lett. \textbf{99}, 190402 (2007).

\bibitem{Holzmann}Markus Holzmann, Werner Krauth, and Martin Naraschewski,
 Phys. Rev. A \textbf{59}, 2956 (1999).

\bibitem{Gerbier2004-experimental-study}F. Gerbier, J. H. Thywissen,
S. Richard, M. Hugbart, P. Bouyer, and A. Aspect, Phys. Rev. A \textbf{70},
013607 (2004).

\bibitem{Opanchuk2013-functional}B. Opanchuk and P. D. Drummond,
 J. Math. Phys. \textbf{54(4)}, 042107 (2013).

\bibitem{Drummond1993}P. D. Drummond and A. D. Hardman, Europhys.
Letters \textbf{21}, 279 (1993).

\bibitem{Steele1998}M. J. Steel, M. K. Olsen, L. I. Plimak, P. D.
Drummond, S. M. Tan, M. J. Collett, D. F. Walls, and R. Graham, Phys.
Rev. A \textbf{58}, 4824 (1998).

\bibitem{laudat-2018-spinsq}T. Laudat, V. Dugrain, T. Mazzoni, M.-Z.
Huang, C. L. Garrido Alzar, A. Sinatra, P. Rosenbusch, and J. Reichel,
 arXiv:1804.07536
\end{thebibliography}
\end{document}